\documentclass[12pt, draftcls, onecolumn]{IEEEtran}

\usepackage{amssymb}
\usepackage{amsmath}
\usepackage{bm}
\usepackage{amsbsy}
\usepackage[mathcal]{eucal}
\usepackage{cite}
\usepackage{graphicx} 
\usepackage{subfigure}
\usepackage{color}
\usepackage[margin=1in]{geometry}

\newtheorem{lemma}{Lemma} 

\newtheorem{proposition}{Proposition}

\begin{document}

\title{\LARGE \vspace{4mm} Difference Antenna Selection and Power Allocation for Wireless Cognitive Systems}  
\author{\authorblockN{Yue Wang, \textit{Member, IEEE}, and Justin P. Coon, \textit{Senior Member, IEEE}}  
\authorblockA{Telecommunication Research Laboratory (TRL), \\Toshiba Research Europe Limited, UK, BS1 4ND \\
Email: {\{yue.wang, justin\}@toshiba-trel.com}}}

\maketitle

\begin{abstract}
In this paper, we propose an antenna selection method in a wireless cognitive radio (CR) system, which we term \emph{difference selection}, whereby a single transmit antenna is selected at the secondary transmitter out of $M$ possible antennas such that the weighted difference between the channel gains of the data link and the interference link is maximized. We analyze the mutual information and the outage probability of the secondary transmission in a CR system with difference antenna selection, and propose a method of optimizing these performance metrics subject to practical constraints on the peak secondary transmit power and the average interference power as seen by the primary receiver. The optimization is performed over two parameters: the peak secondary transmit power and the difference selection weight $\delta\in[0,1]$. Furthermore, we show that the diversity gain of a CR system employing difference selection is an impulsive function of $\delta$, in that a value of $\delta=1$ yields the full diversity order of $M$ and any other value of $\delta$ gives no diversity benefit. Finally, we demonstrate through extensive simulations that, in many cases of interest, difference selection using the optimal values of these two parameters is superior to the so-called \emph{ratio selection} method disclosed in the literature.
\end{abstract}


\newpage

\section{Introduction}\label{introduction}
Cognitive radio (CR) is a promising technology that facilitates efficient use of the radio spectrum. Tremendous efforts have been made to study CR in recent years \cite{Haykin2005, weiss2004spectrum, Akyildiz2008, devroye2006achievable, Sridharan2008, hamdi2009opportunistic, Zhang2008b, Zhang2008a, Zhang2009b}. In particular, considering coexisting CR systems where the secondary user is allowed to transmit as long as it causes a tolerable level of interference to the primary receiver, multiple-antenna techniques that can potentially exploit spatial diversity have been investigated under the context of CR networks  \cite{devroye2006achievable, Sridharan2008, hamdi2009opportunistic, Zhang2008b, Zhang2008a}. To this end, the capacity of the secondary link in multiple-input multiple-output (MIMO) systems was studied in  \cite{devroye2006achievable, Sridharan2008}. Moreover, by employing all of the available antennas simultaneously at the transmitter, it was shown in  \cite{hamdi2009opportunistic} and \cite{Zhang2008b} that the interference caused by the secondary user can be controlled by using beamforming techniques. However, this approach incurs a high computational cost due to the calculations and feedback required to obtain the beamforming vector  \cite{wang2010capacity}. 

Antenna selection is an alternative to full-complexity beamforming that can be used to exploit spatial diversity in an efficient manner  \cite{molisch2004mimo, Sanayei2004}. With transmit antenna selection, instead of transmitting data from all available antennas (say, $M$), a subset of antennas is selected to meet a given criterion, and these antennas are connected to the available radio frequency (RF) chains, which may be fewer in number than the available transmit antennas. A key benefit of antenna selection lies in the reduction in the associated implementation costs  \cite{molisch2004mimo}. Additionally, antenna selection systems achieve the full diversity gain of $M$ \cite{molisch2004mimo}. 

The attractive features of antenna selection have motivated research on this technology within the framework of CR networks. For example, in  \cite{wang2010capacity}, \cite{Zhou2008}. In \cite{wang2010capacity} and \cite{Zhou2008}, a single antenna at the secondary transmitter was selected such that the ratio between the channel gains of the secondary-to-secondary ($s\rightarrow s$) link and the secondary-to-primary ($s\rightarrow p$) link is maximized\footnote{Note that it is assumed that the secondary user must have knowledge of the channel gains for the $s\rightarrow s$ link and the $s\rightarrow p$ link, which can be obtained from feedback in frequency-division duplex (FDD) systems or from channel reciprocity in time-division duplex (TDD) systems.}. In \cite{Zhou2008}, this approach, known as \emph{ratio selection}, was shown to offer a good trade-off between the ergodic capacity of the secondary link and the interference caused to the primary link when a fixed transmit power is used. In \cite{wang2010capacity}, using the same selection method, the ergodic capacity of the $s\rightarrow s$ link was maximized subject to a constraint on the peak power of the interference caused to the primary link (herein referred to as a peak interference power constraint (PIC)). Specifically, it was shown that ratio selection combined with power loading based on instantaneous knowledge of the $s\rightarrow p$ channel yields optimal performance. However, the work in \cite{wang2010capacity} failed to consider the case where the secondary transmission power should also be limited. Such a constraint is usually essential considering the practical power-emission rules such as those stipulated by the Federal Communications Committee (FCC)  \cite{4657371}. 

In this paper, we propose an alternative antenna selection approach for use in CR systems, which we term \emph{difference selection}, where a single antenna is selected according to a weighted difference between the channel gains for the $s\rightarrow s$ and $s\rightarrow p$ links. Based on this selection method, we optimize the mutual information and the outage probability of the secondary link subject to a secondary transmission power constraint and an interference power constraint. Optimization is performed over two parameters: the peak secondary transmit power and the difference selection weight $\delta\in[0,1]$. In contrast to \cite{wang2010capacity} where the \emph{peak} interference power is constrained, we apply an \emph{average} interference power constraint (AIC) at the primary receiver, which is preferable to implementing a PIC in practice in terms of both protecting the quality of the primary link and maximizing the throughput of the secondary link \cite{5199373}. The main contributions of the paper are:
\begin{itemize} 
	\item a difference antenna selection method for CR systems is proposed;
	\item closed-form expressions for the mutual information and the outage probability of the secondary link of a CR system using difference selection are derived as functions of the difference selection weight and the secondary transmit power;
	\item the diversity order of a secondary system employing difference selection is analyzed, and it is shown that this is an impulsive function of $\delta$;
	\item the mutual information and the outage probability of the secondary link are optimized subject to a secondary transmission power constraint and an interference power constraint, where optimization is performed over the weight $\delta$ and the secondary transmit power;
	\item extensive simulation results illustrating the mutual information and the outage probability of CR systems using difference selection and ratio selection are given, and it is shown that difference selection often yields superior performance in practical scenarios.
\end{itemize}

The rest of the paper is organized as follows. In Section II, the model of the CR system with difference selection is described, and the optimization problem is formulated. Section III derives the mutual information and the outage probability of the secondary link with difference selection. The power allocation strategy and the selection weight that optimizes the mutual information and the outage probability are presented in Section IV. Finally, results and comparisons between difference selection and ratio selection are given in Section V, and conclusions are drawn in Section VI.

\emph{Notations:} The probability density function (p.d.f.) and cumulative distribution function (c.d.f.) of a random variable $X$ are denoted as $f_X(x)$ and $F_X(x)$, respectively; the probability of an event $A$ is denoted as $P(A)$, and the conditional probability of an event $A$ given $B$ is denoted as $P(A|B)$; In addition, $\mathsf{E}_1(x)$ denotes the exponential integral function given by $\mathsf{E}_1(x)=\int_x^{\infty}\frac{e^{-u}}{u}du$, and $\mbox{erfc}(x)$ denotes the complementary error function given by $\mbox{erfc}(x)=\frac{2}{\sqrt{\pi}}\int_{x}^{\infty}e^{-t^2}dt$;  $\max\{a_1, \cdots, a_M\}$ and $\min\{a_1, \cdots, a_M\}$ denote the maximum and minimum number among $M$ real numbers $a_1,\cdots, a_M$, respectively; and $F[\cdot]$ and $F(\cdot)$ denote a functional and a function of real arguments, respectively. Finally, $\mathbb{E}$ denotes expectation, and $x\in[a,b]$ denotes that a number $x$ is in the closed interval of $a$ and $b$.  
 
\section{System Model and Problem Formulation}\label{model}
\subsection{Preliminaries and Optimal Problem Formulation}
\begin{figure}[t]
     \centering
     \includegraphics[width=0.5\textwidth]{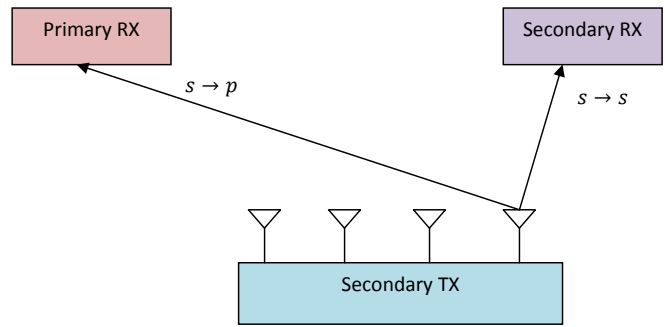}
     \caption{System model of a CR system with antenna selection at the secondary transmitter.}
    \label{fig:system_model}
\end{figure}

We consider a CR system with one primary link and one secondary link. The primary and secondary receivers have one receive antenna, while the secondary transmitter has $M$ transmit antennas. Such a system model is illustrated in Fig.~\ref{fig:system_model}. We consider a co-existing CR system where the secondary user is allowed to transmit subject to a peak transmission power constraint as long as the average interference power caused to the primary system is below a given threshold. We assume that the channel coefficients of the $s\rightarrow s$ link and the $s\rightarrow p$ link fade according to independent Rayleigh distributions. The instantaneous channel gains of the $s\rightarrow s$ link and the $s\rightarrow p$ link corresponding to the $i$th transmit antennas, denoted as $\gamma_{s,i}$ and $\gamma_{p,i}$, respectively, are exponentially distributed random variables. The p.d.f.'s of $\gamma_{s,i}$ and $\gamma_{p,i}$ are given by \cite{proakis-digital}
\begin{equation}\label{fsi}
f_{\gamma_{s,i}}(\gamma) = \frac{1}{\bar{\gamma}_{s}}e^{-\gamma/\bar{\gamma}_{s}}
\end{equation}
and
\begin{equation}\label{fpi}
f_{\gamma_{p,i}}(\gamma) = \frac{1}{\bar{\gamma}_{p}}e^{-\gamma/\bar{\gamma}_{p}}
\end{equation}
respectively, where $\bar{\gamma}_{s}$ and $\bar{\gamma}_p$ are the corresponding average channel gains. 

At each time interval, the secondary transmitter selects one of the $M$ antennas according to a certain criterion to transmit data. Suppose the $\hat{\imath}$th antenna is selected, and the channel gains related to transmission from this secondary transmit antenna to the secondary and primary receivers are denoted by $\gamma_{s,\hat{\imath}}$ and $\gamma_{p,\hat{\imath}}$, respectively. Let $\wp$ be the average interference power limit allowed at the primary receiver, and $P_{\mbox{\tiny{max}}}$ be the maximum allowable transmission power at the secondary transmitter. Ideally, one would determine the optimal antenna selection criterion and the secondary transmission power allocation strategy by solving the following optimization problem 
\begin{equation}\label{problem1}
\begin{array}{ll}
\mbox{optimize} & \mathcal{C}[P_s, f_{\gamma_{s,\hat{\imath}}}(\gamma)]  \\ 
\mbox{s.t.} &   \mathbb{E}\left(P_s\gamma_{p,\hat{\imath}}\right)\leq \wp  \\  
& P_s \leq P_{\mbox{\tiny{max}}}
\end{array}
\end{equation} 
where $\mathcal{C}$ is the objective functional, which can be mutual information or outage probability. Note that the optimization is performed over the p.d.f. of the channel gain between the selected transmit antenna and the secondary receiver $f_{\gamma_{s,\hat{\imath}}}(\gamma)$. In addition, $P_s$ is also a functional dependent on this function. Solving the optimization problem stated above requires knowledge of $f_{\gamma_{s,\hat{\imath}}}(\gamma)$ and $f_{\gamma_{p,\hat{\imath}}}(\gamma)$, which are determined by the selection criterion. Without knowing what this selection criterion is, the optimization problem is intractable. One could perform the optimization over an ensemble of p.d.f.'s, but this is not possible in a practical implementation. Here, we take a two-step approach where we first propose a selection criterion, and then determine the value of $P_s$ that maximizes the mutual information or minimizes the outage probability of the secondary link with antenna selection based on such a criterion. 

\subsection{Alternative Formulation Based on Difference Selection}
Various criteria can be used to select the transmit antenna for the secondary user. For example, one can select the antenna that yields the largest $s\rightarrow s$ channel gain, i.e., select the $\hat{\imath}$th antenna such that $\gamma_{s,\hat{\imath}}=\max\{\gamma_{s,1}, \cdots, \gamma_{s,M}\}$ \cite{Zhou2008}. Alternatively, one can select the antenna that yields the minimum interference to the primary user, i.e., select the $\hat{\imath}$th antenna such that $\gamma_{p,\hat{\imath}}=\min\{\gamma_{p,1}, \cdots, \gamma_{p,M}\}$ \cite{Zhou2008}. These two selection methods are referred to as \emph{maximum data gain selection} and \emph{minimum interference selection}, respectively \cite{Zhou2008}. \emph{Ratio selection}, proposed in \cite{wang2010capacity} and \cite{Zhou2008}, selects the $\hat{\imath}$th antenna such that the ratio of the channel gains of the $s\rightarrow s$ and $s\rightarrow p$ links is maximized, i.e., the $\hat{\imath}$th antenna satisfies $\frac{\gamma_{s,\hat{\imath}}}{\gamma_{p,\hat{\imath}}}=\max\left\{\frac{\gamma_{s,1}}{\gamma_{p,1}}, \cdots,\frac{\gamma_{s,M}}{\gamma_{p,M}}\right\}$. 

In this paper, we propose an alternative antenna selection method for CR systems, which will be shown to be superior to ratio selection in many practical cases. The proposed selection method, referred to as \emph{difference selection}, selects the antenna at the secondary transmitter such that the weighted difference between the channel gains for the $s\rightarrow s$ link and the $s\rightarrow p$ link is maximized. Denote the selection weight as $\delta\in[0,1]$. Difference selection selects the $\hat{\imath}$th antenna such that $Z_{\hat{\imath}}=\max\left\{Z_1, \cdots, Z_M\right\}$, where $Z_i$ ($i=1,\cdots, M$) is given by
\begin{equation}\label{Zi}
Z_i = \delta\gamma_{s,i} - (1-\delta)\gamma_{p,i}.
\end{equation}
Note that difference selection becomes minimum interference selection when $\delta=0$, and maximum data gain selection when $\delta=1$. 

With difference selection, the mutual information and outage probability for the secondary link are dependent upon $\delta$. In the following, we formulate optimization problems to jointly optimize $\delta$ and the secondary transmission power, such that the mutual information and outage probability are optimized subject to constraints on the peak transmission power for the secondary user and the average interference power affecting the primary receiver. 

Now, suppose difference selection selects the $\hat{\imath}$th antenna to transmit data at a given time slot. The optimization problem is formulated as follows: 
\begin{equation}\label{problem2}
\begin{array}{ll}
\mbox{optimize} & \mathcal{C}(P_s, \delta)  \\ 
\mbox{s.t.} &   \mathbb{E}\left(P_s\gamma_{p,\hat{\imath}}\right)\leq \wp  \\  
& P_s \leq P_{\mbox{\tiny{max}}}
\end{array}
\end{equation} 
where $\mathcal{C}(P_s, \delta)$ is the objective function, which can be mutual information or outage probability, and $P_s$ is the secondary transmission power that is to be determined, which is a function of the instantaneous channel gains $\gamma_{s,\hat{\imath}}$ and $\gamma_{p,\hat{\imath}}$, and thus a function of $\delta$. Note that in the problem considered here, the optimization is performed over two variables: the secondary transmission power $P_s$ and and selection weight $\delta$. Therefore, in contrast to the optimization problem given in (\ref{problem1}), the objective function $\mathcal{C}$ in this case is a function, not a functional, and the optimization process can be significantly simplified. In fact, for a given $\delta$, the optimal power loading strategy for such an optimization problem is addressed in \cite{Zhang2008a}, where the secondary transmission power $P_s$ is defined as a function of the instantaneous channel gains $\gamma_{s,\hat{\imath}}$ and $\gamma_{p,\hat{\imath}}$, and is given by 
\begin{equation}\label{optp1}
P_s(\gamma_{s,\hat{\imath}}, \gamma_{p,\hat{\imath}}) = \left\{\begin{array}{ll}
0, & \gamma_{p,\hat{\imath}} \geq \frac{\log_2e}{\lambda N_0}\gamma_{s,\hat{\imath}} \\
\frac{\log_2e}{\lambda \gamma_{p,\hat{\imath}}}-\frac{N_0}{\gamma_{s,\hat{\imath}}}, & \frac{\log_2e}{\lambda N_0}\gamma_{s,\hat{\imath}}>\gamma_{p,\hat{\imath}}>\frac{\log_2e}{\lambda\left(P_{\mbox{\tiny{max}}}+\frac{N_0}{\gamma_{s,\hat{\imath}}}\right)}\\
P_{\mbox{\tiny{max}}}, & \gamma_{p,\hat{\imath}} \leq \frac{\log_2e}{\lambda\left(P_{\mbox{\tiny{max}}}+\frac{N_0}{\gamma_{s,\hat{\imath}}}\right)}\end{array}\right.
\end{equation}
where $N_0$ is the noise power, and $\lambda$ is determined by substituting (\ref{optp1}) into the average interference power constraint given in (\ref{problem2}), i.e., $\lambda$ is defined implicitly by the equation
\begin{equation}
\int_{0}^{\infty}\int_{0}^{\infty}P_s(\gamma_{s,\hat{\imath}}, \gamma_{p,\hat{\imath}})\gamma_{p,\hat{\imath}}f_{\gamma_{s,\hat{\imath}}, \gamma_{p,\hat{\imath}}}(\gamma_{s,\hat{\imath}}, \gamma_{p,\hat{\imath}})d\gamma_{s,\hat{\imath}}d\gamma_{p,\hat{\imath}} = \wp
\end{equation} 
where $f_{\gamma_{s,\hat{\imath}}, \gamma_{p,\hat{\imath}}}(\gamma_{s,\hat{\imath}}, \gamma_{p,\hat{\imath}})$ is the joint p.d.f. of $\gamma_{s,\hat{\imath}}$ and $\gamma_{p,\hat{\imath}}$. These two random variables are dependent due to the selection process. The key to solving the optimization problem therefore lies in the derivation of $f_{\gamma_{s,\hat{\imath}}, \gamma_{p,\hat{\imath}}}(\gamma_{s,\hat{\imath}}, \gamma_{p,\hat{\imath}})$. Unfortunately, the expression for this joint p.d.f. can be complicated, thus making it difficult to determine the optimal transmit power analytically. 

\subsection{Practical Formulation}
The problem given in (\ref{problem1}) can be made practical by considering power allocation based on channel statistics rather than instantaneous channel knowledge. In such a case, the optimization problem becomes 
\begin{equation}\label{problem3}
\begin{array}{ll}
\mbox{optimize} & \mathcal{C}(P_s, \delta)  \\  
\mbox{s.t.} &   P_s\mathbb{E}\left(\gamma_{p,\hat{\imath}}\right)\leq \wp  \\   
& P_s \leq P_{\mbox{\tiny{max}}} 
\end{array}
\end{equation}
where, in contrast to (\ref{problem2}), $P_s$ is taken out of the expectation function in (\ref{problem3}) because it is not a function of the instantaneous channel gains $\gamma_{p,\hat{\imath}}$. With the optimization problem considered above, to determine the optimal transmission power, only the mean of $\gamma_{p,\hat{\imath}}$ is required. Even so, when ratio selection is considered, the calculation of this mean does not have a closed form. Fortunately, by using difference selection at the secondary transmitter, a closed-form expression for the mean of the $s\rightarrow p$ link gain can be obtained, based on which the mutual information and the outage probability can be optimized. Not only does the use of difference selection facilitate the mathematical tractability of the optimization problem, but we further show that, with the practical peak secondary transmission power and average interference power constraints, difference selection using optimal values of $P_s$ and $\delta$ is, in many cases of interest, superior to ratio selection with respect to performance. The advantages of using difference selection will be detailed later. In the following, we first provide an analysis of the mutual information and the outage probability of the secondary link for CR systems with difference selection, and then solve for selection weight $\delta$ and the secondary transmission power $P_s$ that optimize these objective functions. 
 
\section{Mutual Information and Outage Probability Analysis}\label{capacities} 
We now derive the mutual information and the outage probability for the secondary transmission in CR systems with difference antenna selection. In the ensuing analysis, we make use of the following lemma, the proof of which is given in Appendix \ref{appendix1}. 
\begin{lemma}\label{lemma1}
The c.d.f. of the $s\rightarrow s$ channel gain  due to the selection of the $\hat{\imath}$th antenna using difference antenna selection, denoted as $\gamma_{s,\hat{\imath}}$, is given by
\begin{eqnarray}\label{Fhatxxa}
F_{\gamma_{s,\hat{\imath}}}(x) = \left \{\begin{array}{lll}                     
   \sum_{k=0}^{M-1}\rho_k(x)-\frac{\varpi_p^M}{\bar{\gamma}^{M}}e^{-\frac{\delta\bar{\gamma}}{\varpi_s\varpi_p}x}+\left(1-\frac{\varpi_s}{\bar{\gamma}}e^{-\frac{x}{\bar{\gamma}_s}}\right)^M&,& \varpi_s\neq g\varpi_p  \\
  \sum_{k=0\atop k\neq g}^{M-1}\rho_k(x)-\frac{\varpi_p^M}{\bar{\gamma}^{M}}e^{-\frac{\delta\bar{\gamma}}{\varpi_s\varpi_p}x}+\left(1-\frac{\varpi_s}{\bar{\gamma}}e^{-\frac{x}{\bar{\gamma}_s}}\right)^M-\frac{M}{\bar{\gamma}}\delta x\mu_g(x)&,& \varpi_s=g\varpi_p 
    \end{array}\right. 
\end{eqnarray} 
where  
\begin{equation}
\rho_k(x)= M\left(M-1 \atop k\right)\frac{\left(-\varpi_s\right)^{k+1}\varpi_p}{\bar{\gamma}^{k+1}\left(\varpi_s-k\varpi_p\right)}\left(e^{-\frac{(k+1)}{\bar{\gamma}_s}x}-\mu_k(x)\right)
\end{equation}
\begin{equation}
\mu_g(x)= \left(M-1\atop g\right)\frac{(-\varpi_s)^g}{\bar{\gamma}^g}e^{-\frac{g+1}{\bar{\gamma}_s}x}.
\end{equation}
In the equations above, $\varpi_s=(1-\delta)\bar{\gamma}_s$, $\varpi_p=\delta\bar{\gamma}_p$, $\bar{\gamma}=\varpi_s+\varpi_p$, which are all functions of $\delta$, and $g\in[0,M-1]$ is an integer.
\end{lemma} 

\subsection{Mutual Information}
Assuming Gaussian signaling is employed, the maximum mutual information\footnote{In the unit of Bits/Sec/Hz.} of the secondary link for CR systems with difference selection is given by
\begin{equation}\label{ergodic}
R_{\mbox{\tiny{max}}} = \log_2e\max_{P_s,\delta}R(P_s,\delta)
\end{equation}
where $P_s$ and $\delta$ are subject to the constraints given in (\ref{problem3}), and
\begin{equation}\label{R}
R(P_s,\delta)= \int_{0}^{\infty}\log\left(1+\frac{P_sx}{N_0}\right)dF_{\gamma_{s,\hat{\imath}}}(x).
\end{equation} 
In the equation above, $N_0$ is the noise power and $F_{\gamma_{s,\hat{\imath}}}(\cdot)$ is the c.d.f. of $\gamma_{s,\hat{\imath}}$, given in Lemma 1. Substituting (\ref{Fhatxxa}) into (\ref{R}), we have  
\begin{equation}\label{Rpsdelta}
R(P_s,\delta) = \left\{\begin{array}{ll}\sum_{k=0}^{M-1}(\Psi_k(P_s,\delta)-\Phi_k(P_s,\delta))+\Upsilon(P_s,\delta), & \varpi_s\neq g\varpi_p \\
\sum_{k=0\atop k\neq g}^{M-1}\Psi_k(P_s,\delta)-\sum_{k=0}^{M-1}\Phi_k(P_s,\delta)+\Upsilon(P_s,\delta)+\Theta_g(P_s,\delta), & \varpi_s=g\varpi_p\end{array}\right.
\end{equation} 
where
\begin{equation}
\Psi_k(\delta, P_s)=M\left(M-1 \atop k\right)\frac{\left(-\varpi_s\right)^{k+1}\varpi_p}{\bar{\gamma}^{k+1}\left(\varpi_s-k\varpi_p\right)}\left(-e^{\frac{(k+1)N_0}{\bar{\gamma}_sP_s}}\mathsf{E}_1\left(\frac{(k+1)N_0}{\bar{\gamma}_sP_s}\right)+e^{\frac{\delta \bar{\gamma} N_0}{\varpi_s\varpi_p P_s}}\mathsf{E}_1\left(\frac{\delta \bar{\gamma}N_0}{\varpi_s\varpi_pP_s}\right)\right)\nonumber \\
\end{equation}
\begin{equation}
\Upsilon(P_s,\delta) = \frac{\varpi_p^M}{\bar{\gamma}^M}e^{\frac{\delta\bar{\gamma}N_0}{\varpi_s\varpi_pP_s}}\mathsf{E}_1\left(\frac{\delta\bar{\gamma}N_0}{\varpi_s\varpi_pP_s}\right)\nonumber 
\end{equation}
\begin{equation}
\Phi_k(\delta, P_s)=M\left({M-1} \atop k\right)\left(-\frac{\varpi_s}{\bar{\gamma}}\right)^{k+1}\frac{1}{k+1}e^{\frac{(k+1)N_0}{\bar{\gamma}_sP_s}}\mathsf{E}_1\left(\frac{(k+1)N_0}{\bar{\gamma}_sP_s}\right)\nonumber
\end{equation}
and
\begin{equation}
\Theta_g(\delta, P_s)= -\left(M\atop g+1\right)\left(\frac{-\varpi_s}{\bar{\gamma}}\right)^{g+1}+M\left(M-1\atop g\right)\left(\frac{-\varpi_s}{\bar{\gamma}}\right)^{g+1}\frac{N_0}{\bar{\gamma}_sP_s}e^{\frac{(g+1)N_0}{\bar{\gamma}_sP_s}}\mathsf{E}_1\left(\frac{(g+1)N_0}{\bar{\gamma}_sP_s}\right)\nonumber. 
\end{equation}  

From (\ref{Rpsdelta}), it is known that $R$ is a function of the secondary transmission power $P_s$ and the selection weight $\delta$. The method of determining $P_s$ and $\delta$ to maximize $R$ will be discussed later. For now, assume $P_s$ is given. It can be verified that, in such a case, $R$ is a monotonically increasing function of $\delta$. In particular, the maximum $R$ is achieved when $\delta=1$, which is given by
\begin{equation}\label{R1}
\lim_{\delta\rightarrow 1} R(P_s,\delta) = \sum_{k=0}^{M-1}\left(M \atop k+1\right)(-1)^ke^{\frac{(k+1)N_0}{\bar{\gamma}_sP_s}}\mathsf{E}_1\left(\frac{(k+1)N_0}{\bar{\gamma}_sP_s}\right). 
\end{equation}
The expression of $R$ when $\delta \rightarrow 1$ coincides with that given for a conventional non-CR system with selection combining (c.f. (44) in \cite{alouini1999capacity}). This is intuitively correct because when $\delta=1$, for CR systems with difference selection, data from the secondary user is transmitted as if the primary user does not exist, and difference antenna selection essentially becomes the maximum data gain selection.  

Similarly, when $\delta=0$, from the secondary user's perspective, antenna diversity is not exploited, and the data is effectively transmitted through a Rayleigh fading channel from a single (random) antenna. In such a case, $R$ reaches its minimum and it remains the same regardless of the number of transmit antennas employed at the secondary transmitter. Following (\ref{Rpsdelta}), one can verify that when $\delta=0$, $R$ indeed becomes the capacity of a single antenna transmission through a Rayleigh fading channel (c.f. (34) in \cite{alouini1999capacity}), which is given by  
\begin{equation}\label{R2}
\lim_{\delta\rightarrow 0}R(P_s, \delta) =e^{\frac{N_0}{\bar{\gamma}_sP_s}}\mathsf{E}_1\left(\frac{N_0}{\bar{\gamma}_sP_s}\right).
\end{equation} 

%For a given $\gamma$, $R$ of the secondary link as a function of $\delta$ is shown in Fig.~\ref{ecapacity2}. In addition, asymptotic $R$ of the secondary link using different number of antennas is shown in Fig.~\ref{ecapacity}. 
%
%\begin{figure}[t]
%     \centering
%     \includegraphics[width=\textwidth]{tmp1.eps}
%     \caption{Ergodic Capacity}
%    \label{ecapacity}
%\end{figure}
%
%
%\begin{figure}[t]
%     \centering
%     \includegraphics[width=\textwidth]{tmp2.eps}
%     \caption{Ergodic Capacity with different $\delta$}
%    \label{ecapacity2}
%\end{figure}
%

\subsection{Outage Probability Analysis}
The outage probability of the secondary link for a CR system with difference selection and a given outage transmission rate $r_0$ is given by \cite{tse2005fundamentals}
\begin{equation}\label{outage}
P_{\mbox{\tiny{out}}} = \min_{P_s,\delta}P(P_s,\delta)
\end{equation}
where $P_s$ and $\delta$ are subject to the constraints given in (\ref{problem3}), and $P(P_s, \delta)$ is the probability that the rate of the secondary transmission is smaller than or equal to a given $r_0$, given by \cite{tse2005fundamentals}
\begin{equation}\label{Pr1}
P(P_s,\delta)= F_{\gamma_{s,\hat{\imath}}}\left(\frac{(2^{r_0}-1)N_0}{P_s}\right)
\end{equation}
and $F_{\gamma_{s,\hat{\imath}}}(\cdot)$ is given in Lemma 1. 

\subsubsection{Asymptotic Analysis}
It is interesting to study the diversity order of the secondary system to obtain insight into the secondary link performance at high SNR. Following (\ref{Pr1}), when $P_s\rightarrow \infty$, the diversity order can be gleaned from the expression for $F_{\gamma_{s,\hat{\imath}}}(x)$ when $x\rightarrow 0$. Applying the first order Taylor expansion of exponential functions $e^x=1+x+O(x^2)$ to (\ref{Fhatxxa}), one can confirm that   
\begin{equation}\label{Fhatxx0}
F_{\gamma_{s,\hat{\imath}}}(x\rightarrow 0)  = \left \{\begin{array}{lll}      
    \left(\frac{x}{\bar{\gamma}_s}\right)^M +O(x^{M+1}) &,&\delta =1 \\               
    \frac{\varpi_p^{M-1}}{\bar{\gamma}^{M-1}\bar{\gamma}_s}x + O(x^2)&,& \delta\neq 1, \varpi_s\neq g\varpi_p  \\
    \frac{1}{(g+1)^{M-1}\bar{\gamma}_s}x+O(x^2)&,& \delta\neq 1, \varpi_s=g\varpi_p 
    \end{array}\right..
\end{equation} 
Recall that in the equation above, $\varpi_s=\delta\bar{\gamma}_s$, $\varpi_p=(1-\delta)\bar{\gamma}_p$, and $\bar{\gamma}=\varpi_s+\varpi_p$, which are all functions of $\delta$, and $g$ is an integer where $g\in[0,M-1]$. 

Equation (\ref{Fhatxx0}) indicates that the secondary transmission of a CR system with difference antenna selection achieves a diversity order of $M$ when $\delta=1$. Indeed, when $\delta=1$, a CR system with difference antenna selection is essentially a conventional antenna selection system from the secondary user's point of view. Such a conventional system has been investigated in \cite{chen2005analysis}, and our result for $\delta=1$ agrees with the expression for the outage probability given there (c.f. (11) in \cite{chen2005analysis}). For any other $\delta\neq 1$, however, the secondary user only achieves a diversity order of $1$. The same results on the diversity order of the secondary link with difference selection are also verified in \cite{CRconference} via bit error rate (BER) analysis. 

It is worth considering the implications of the impulse-like nature of the diversity order detailed in the calculations above.  Effectively, this result suggests that any consideration made to reduce the interference to a primary user via antenna selection negates all beneficial effects of employing multiple antennas in the secondary link at high SNR.  Therefore, if the secondary transmit power is constant, there is little purpose in setting $\delta$ close to, but strictly less than one, since all diversity gain is surrendered  in doing so and only a minor emphasis is placed on reducing interference to the primary user.  However, if the secondary transmit power is allowed to vary with changing channel statistics (i.e., mean channel gains), then there turns out to be an intricate relationship between the optimal power level and the weight $\delta$ at finite SNR.  This relationship and how it can be exploited to optimize the mutual information and the outage probability is considered in the next section.

\section{Power Allocation Strategy and Optimal Selection Weight}\label{optimal} 
Having obtained the expressions for mutual information and outage probability given in (\ref{Rpsdelta}) and (\ref{Pr1}), respectively, we are in a position to determine the optimal selection weight and power allocation strategy. First, we introduce the following lemma:
\begin{lemma}\label{lemma2}
The average $s\rightarrow p$ link channel gain due to the selection of the $\hat{\imath}$th antenna using difference antenna selection, denoted as $\mathbb{E}\left(\gamma_{p,\hat{\imath}}\right)$, is given by
\begin{equation}\label{Eyy}
\mathbb{E}\left(\gamma_{p,\hat{\imath}}\right)  =\frac{M(\varpi_s+\varpi_p)^{M-1}\varpi_s+\varpi_p^M}{M(\varpi_s+\varpi_p)^M}\bar{\gamma}_p
\end{equation}
where $M$ is the number of secondary transmit antennas, $\varpi_s=(1-\delta)\bar{\gamma}_s$, and $\varpi_p=\delta\bar{\gamma}_p$. 
\end{lemma}

The proof of Lemma 2 is given in Appendix \ref{appendix4}. Applying (\ref{Eyy}), the left hand side of the first inequality constraint given in (\ref{problem3}) can be rewritten as
\begin{equation}\label{Ipsalpha}
I(P_s, \delta) = P_s\mathbb{E}\left(\gamma_{p,\hat{\imath}}\right) =P_s\frac{M(\alpha+1)^{M-1}\alpha+1}{M(\alpha+1)^M}\bar{\gamma}_p
\end{equation}
where $\alpha=\frac{\varpi_s}{\varpi_p}=\frac{\delta\bar{\gamma}_s}{(1-\delta)\bar{\gamma}_p}$, which is a function of $\delta$. One can confirm that $I(P_s,\delta)$ is monotonically increasing in $\alpha\in (0,+\infty)$ by verifying that the first derivative of $\alpha$ is greater than 0. Therefore, $I$ reaches its maximum of $P_s\bar{\gamma}_p$ when $\alpha\rightarrow \infty$, or equivalently when $\delta=1$. In addition, it reaches its minimum of $P_s\bar{\gamma}_p/M$ when $\alpha=0$, or equivalently when $\delta=0$. In other words, using the proposed difference selection method, the average interference power is always guaranteed to be in a range of $I\in[P_s\bar{\gamma}_p/M, P_s\bar{\gamma}_p]$. In particular, compared to the maximum data gain selection ($\delta=1$), the interference power can be reduced by a factor of $M$ when minimum interference power selection ($\delta=0$) is performed for antenna selection. 

Having noticed that, for a given $\delta$, $R(P_s)$ and $P(P_s)$ are monotonically increasing and decreasing functions of $P_s$, respectively, the maximum of $R$ and the minimum of $P$ are reached when $P_s$ takes the maximum possible value. In the case where $\bar{\gamma}_p$ is sufficiently small\footnote{For example, when the secondary transmitter is far away from the primary receiver, or there is obstacle material that yields deep fading channels between the two.} such that $P_{\mbox{\tiny{max}}}\bar{\gamma}_p\leq \wp$, the interference constraint is inactive and one should let $P_s=P_{\mbox{\tiny{max}}}$. When $P_{\mbox{\tiny{max}}}\bar{\gamma}_p>\wp$, the average interference constraint is active, and the optimal secondary transmission power is the minimum between $P_{\mbox{\tiny{max}}}$ and the transmission power that satisfies the interference constraint with equality. Applying (\ref{Ipsalpha}), we have the following power allocation strategy:
\begin{proposition}
In a CR system with difference antenna selection, for a given selection weight $\delta$, the optimal secondary transmission power $P_s^*$ that maximizes the mutual information and minimizes the outage probability, subject to the constraints that 1) the average interference power is at most $\wp$, and 2) the peak secondary transmission power is at most $P_{\tiny{\mbox{max}}}$, is given by
\begin{equation}\label{Psdelta}
P_s^* = \left\{\begin{array}{lll} \min\left\{\frac{\wp M(\alpha+1)^M}{\left(M(\alpha+1)^{M-1}\alpha+1\right)\bar{\gamma}_p}, P_{\mbox{\tiny{max}}}\right\}&,& \bar{\gamma}_p\geq \wp/P_{\mbox{\tiny{max}}}\\
P_{\mbox{\tiny{max}}}&, & \bar{\gamma}_p<\wp/P_{\mbox{\tiny{max}}} \end{array}\right.
\end{equation}
where $\bar{\gamma}_s$ and $\bar{\gamma}_p$ are the average channel gains of the $s\rightarrow s$ and $s\rightarrow p$ links, respectively, $M$ is the number of antennas, and $\alpha=\frac{\delta\bar{\gamma}_s}{(1-\delta)\bar{\gamma}_p}$.
\end{proposition}

Note that $P_s^*$ is a function of $\delta$ because $\alpha$ is a function of $\delta$. Substituting (\ref{Psdelta}) into (\ref{Rpsdelta}) and (\ref{Pr1}), and noticing the fact that $R(\delta)$ and $P(\delta)$ are concave and convex functions of $\delta$, respectively, the optimal $\delta$ that maximizes mutual information or minimizes outage probability can be determined by using existing numerical techniques (see, e.g., \cite{press2007numerical}).   

%\section{BER Analysis}\label{ber}
%\emph{\color{blue}{This section derives the BER of the secondary system based on difference selection. }}

\section{Results and Discussions}\label{simulations}
In this section, we present simulation results for the mutual information and the outage probability of the secondary link for a CR system using difference antenna selection, and compare them with a similar system using ratio selection. The results are generated by using Rayleigh fading channels for both the $s \rightarrow s$ link and the $s\rightarrow p$ link, with the mean of the channel gains being $\bar{\gamma}_s$ and $\bar{\gamma}_p$, respectively. Unless otherwise specified, we assumed that the interference power threshold $\wp=1$, which is the same for both AIC and PIC, and is equivalent to the noise power \cite{wang2010capacity}, \cite{5199373}. In addition, all the results for outage probability are generated by assuming  $r_0=1$. By way of example, we used $M=2$ or $4$ transmit antennas at the secondary transmitter to study mutual information and outage probability. Using larger numbers of transmit antennas leads to results and trends that are similar to those shown here. 

\begin{figure}[t]
     \centering
     \includegraphics[width=\textwidth]{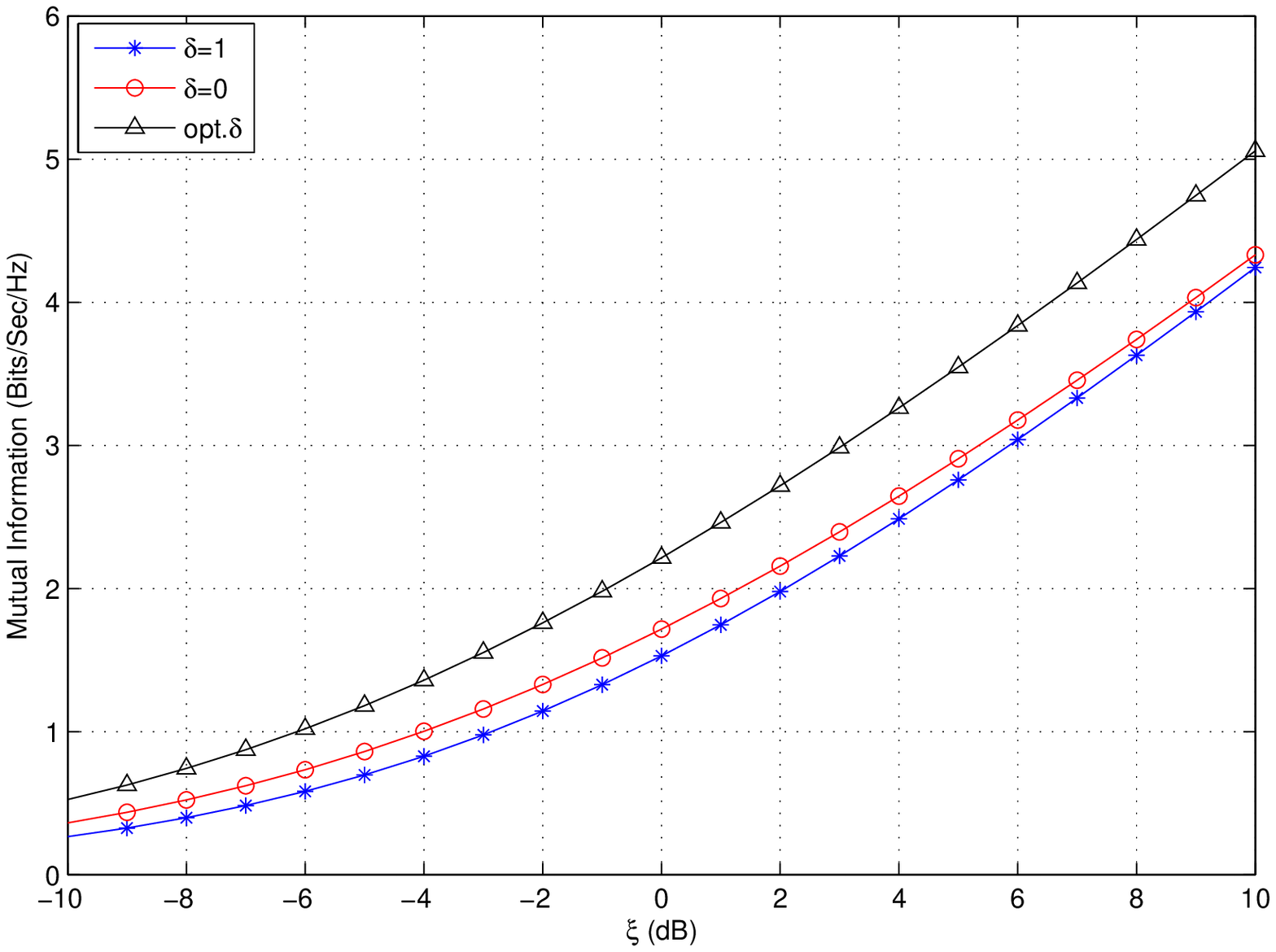}
     \caption{Mutual information of the $s\rightarrow s$ link as a function of $\xi=\bar{\gamma}_s/\bar{\gamma}_p$, where different $\delta$'s are used to select a single antenna among $M=4$ antennas at the secondary transmitter based on difference antenna selection.}
    \label{mutualinf1}
\end{figure}

\begin{figure}[t]
     \centering
     \includegraphics[width=\textwidth]{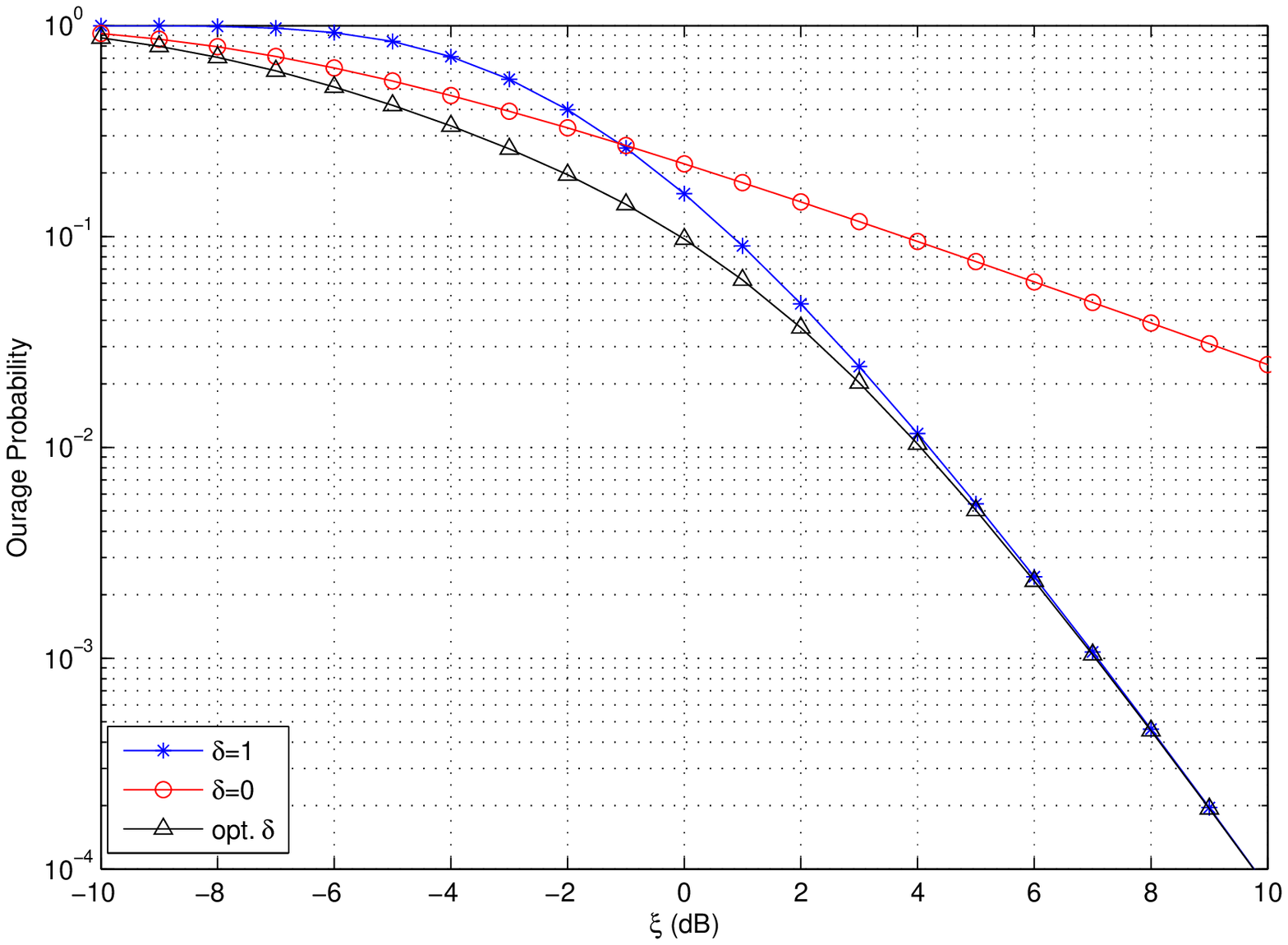}
     \caption{Outage probability of the $s\rightarrow s$ link as a function of $\xi=\bar{\gamma}_s/\bar{\gamma}_p$, where different $\delta$'s are used to select a single antenna among $M=4$ antennas at the secondary transmitter based on difference antenna selection.}
    \label{outageprob1}
\end{figure}

We first show the results of mutual information and outage probability for the secondary link of a CR system with difference antenna selection using the optimal $\delta$, where $M=4$ antennas are used at the secondary transmitter. These results are presented in Figs. \ref{mutualinf1} and \ref{outageprob1}, where the x-axis is $\xi=\bar{\gamma}_s/\bar{\gamma}_p$, and the maximum allowed transmission power is $P_{\tiny{\mbox{max}}}=10$ dB, to follow the convention of \cite{wang2010capacity}. Mutual information and outage probability with the optimal $\delta$ are obtained by first substituting (\ref{Psdelta}) into (\ref{Rpsdelta}) and (\ref{Pr1}), solving for the optimal $\delta$, and then applying this $\delta$ to the respective objective functions. Note that the optimal $\delta$ is also a function of $\xi$. For $\delta=1$ and $\delta=0$, the secondary transmission power is $P_s=\min\left\{\frac{\wp}{\bar{\gamma}_p}, P_{\tiny{\mbox{max}}}\right\}$ and $P_s =\min\left\{\frac{M\wp}{\bar{\gamma}_p}, P_{\tiny{\mbox{max}}}\right\}$, respectively, as given by (\ref{Psdelta}). It is shown in both figures that, comparing to mutual information and outage probability for the secondary link with difference antenna selection using $\delta=1$ and $\delta=0$, significant gains can be observed when the optimal $\delta$ is used. 

Next, we present the mutual information and the outage probability of CR systems with difference selection by using the optimal $\delta$ and the power allocation strategy proposed in this paper, and compare the results with those obtained by using ratio selection with PIC or AIC and different power allocation strategies. In all figures that follow, we assumed the number of antennas at the secondary transmitter is $M=2$. For difference selection, we considered an average interference constraint (DS-AIC), where the results are obtained by using the power allocation strategy and the optimal $\delta$ presented in this paper. For ratio selection, when AIC is considered (RS-AIC), the power allocation strategy is the same as that given in this paper except that the mean of the $s\rightarrow p$ channel is simulated by observing a sufficiently large number of channels since it cannot be calculated in closed form. For ratio selection with PIC and power allocation based on instantaneous channel knowledge (RS-PIC), the results are obtained by using a secondary transmission power of $P_s^*=\min\left\{P_{\mbox{\tiny{max}}}, \frac{\wp}{\gamma_{p,\hat{\imath}}}\right\}$. When $P_{\mbox{\tiny{max}}}\rightarrow \infty$, the transmission power becomes $P_s^*=\frac{\wp}{\gamma_{p,\hat{\imath}}}$, which is essentially the power allocation method given in \cite{wang2010capacity}. We must emphasize that in practice, a peak transmission power constraint shall be applied to the secondary transmitter. Therefore, the results for systems without a peak transmission power constraint shown here are impractical, and are used as benchmarks only. 

\begin{figure}[t]
     \centering
     \includegraphics[width=\textwidth]{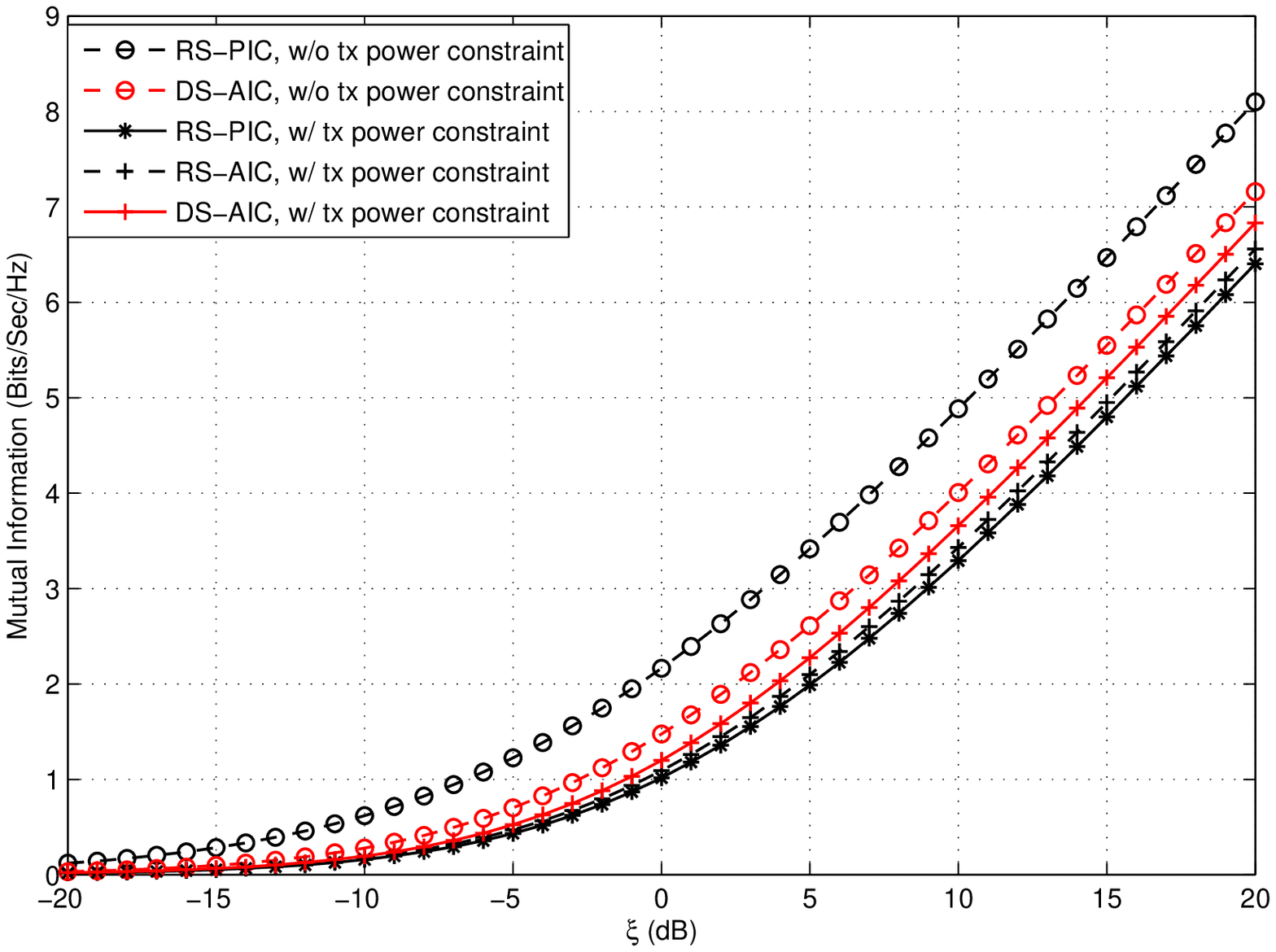}
     \caption{Comparison of mutual information between ratio selection and difference selection with various transmission and interference power constraints as a function of $\xi=\bar{\gamma}_s/\bar{\gamma}_p$, where $M=2$ antennas at the secondary transmitter are used, and $P_{\tiny{\mbox{max}}}=0$ dB.}
    \label{mutualinf2}
\end{figure}

\begin{figure}[t]
     \centering
     \includegraphics[width=\textwidth]{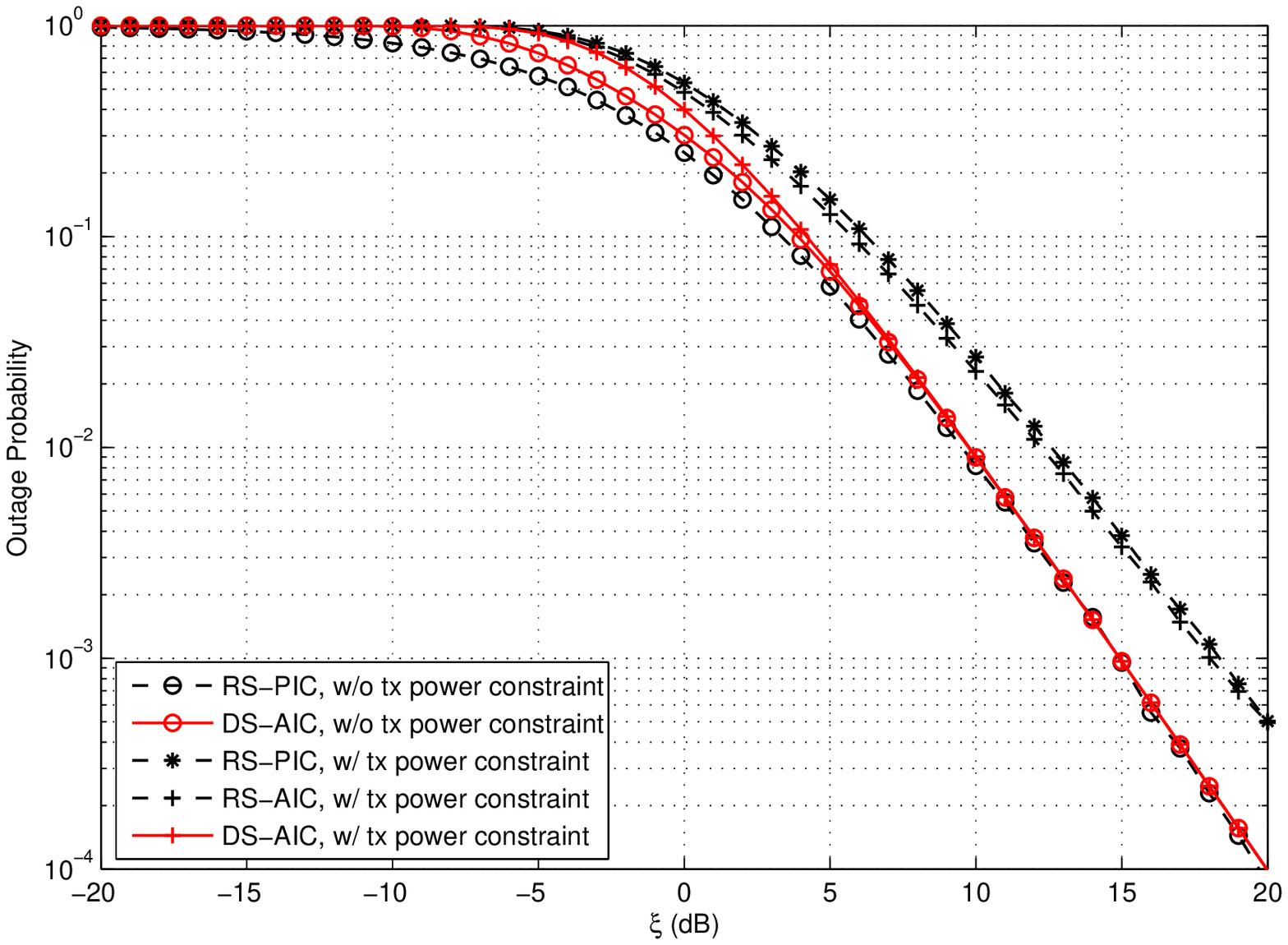}
     \caption{Comparison of outage probability between ratio selection and difference selection with various transmission and interference power constraints as a function of $\xi=\bar{\gamma}_s/\bar{\gamma}_p$, where $M=2$ antennas at the secondary transmitter are used, and $P_{\tiny{\mbox{max}}}=0$ dB.}
    \label{outageprob2}
\end{figure}

We first show, in Figs.~\ref{mutualinf2} and \ref{outageprob2}, the results for mutual information and outage probability as a function of $\xi=\bar{\gamma}_s/\bar{\gamma}_p$. It is observed from Figs.~\ref{mutualinf2} and \ref{outageprob2} that, for the impractical case where no peak transmission power constraint is applied, i.e., when $P_{\tiny{\mbox{max}}}=+\infty$, RS-PIC is indeed optimal in the sense that it provides the maximum mutual information and the minimum outage probability among the systems considered. However, in practice, when the secondary transmission power is constrained, the performance of the secondary link with RS-PIC degrades considerably. For example, it is observed from Fig.~\ref{outageprob2} that about a $3$ dB degradation occurs at an outage probability of $10^{-2}$ when a stringent transmission power constraint of $P_{\tiny{\mbox{max}}}=0$ dB is applied. One can consider AIC and apply the same power allocation strategy described in this paper to ratio selection. In such a case, RS-AIC yields a slightly better performance in the secondary link compared to the case where RS-PIC is employed. This results from the fact that AIC is a more relaxed constraint compared to PIC from the perspective of secondary transmission. A comparison between the performance of ratio selection and difference selection shows that difference selection yields inferior mutual information and outage probability without a peak transmission power constraint. However, considering the practical case when such a constraint is applied, performance of a CR system employing DS-AIC significantly outperforms  systems using RS-AIC or RS-PIC.

\begin{figure}[t]
     \centering
     \includegraphics[width=\textwidth]{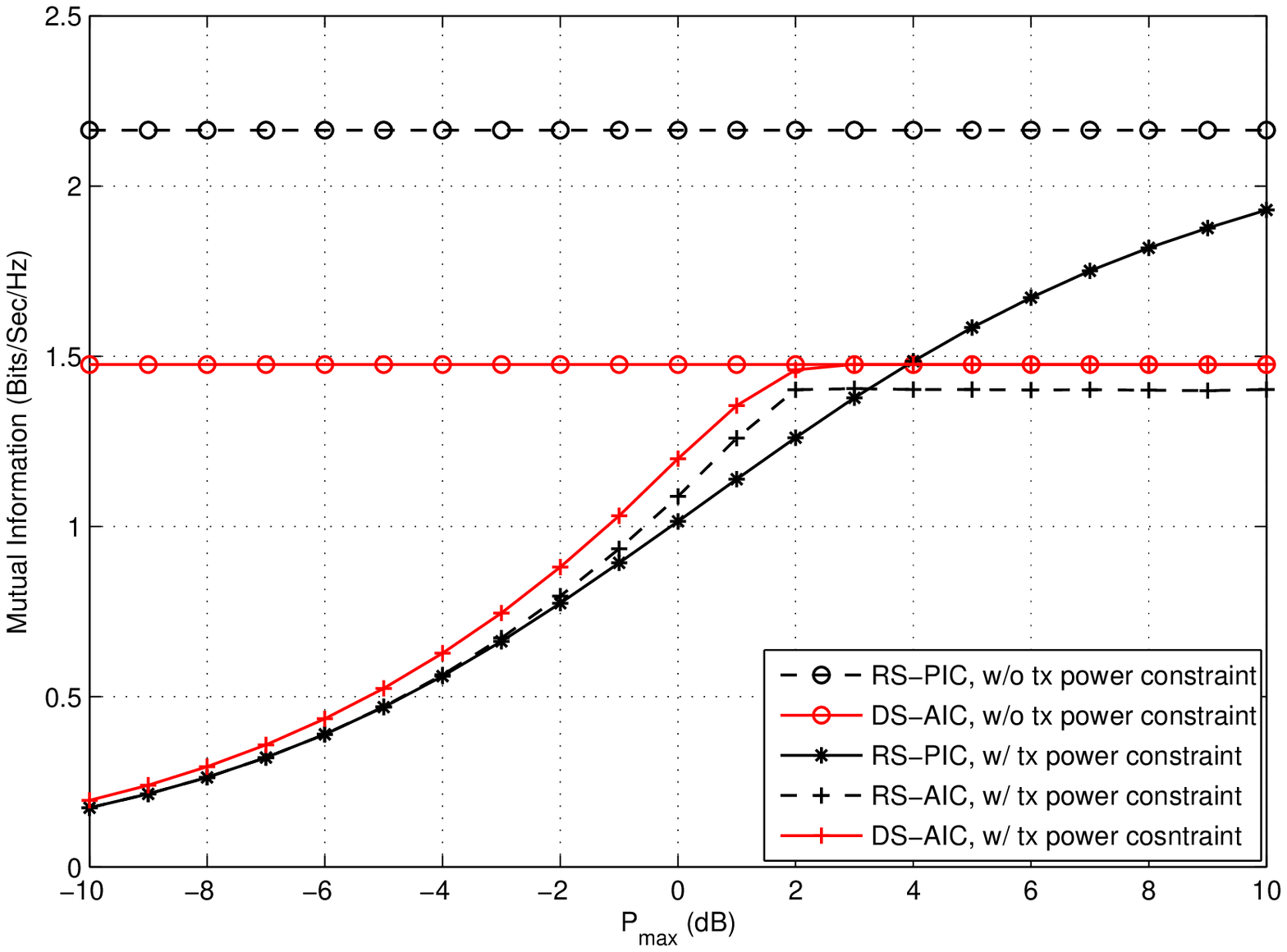}
     \caption{Comparison of mutual information between ratio selection and difference selection with various transmission and interference power constraints as a function of maximum allowable secondary transmission power $P_{\tiny{\mbox{max}}}$, where $M=2$ antennas at the secondary transmitter are used, and $\bar{\gamma}_s=\bar{\gamma}_p=1$.}
    \label{Pmax1}
\end{figure}

\begin{figure}[t]
     \centering
     \includegraphics[width=\textwidth]{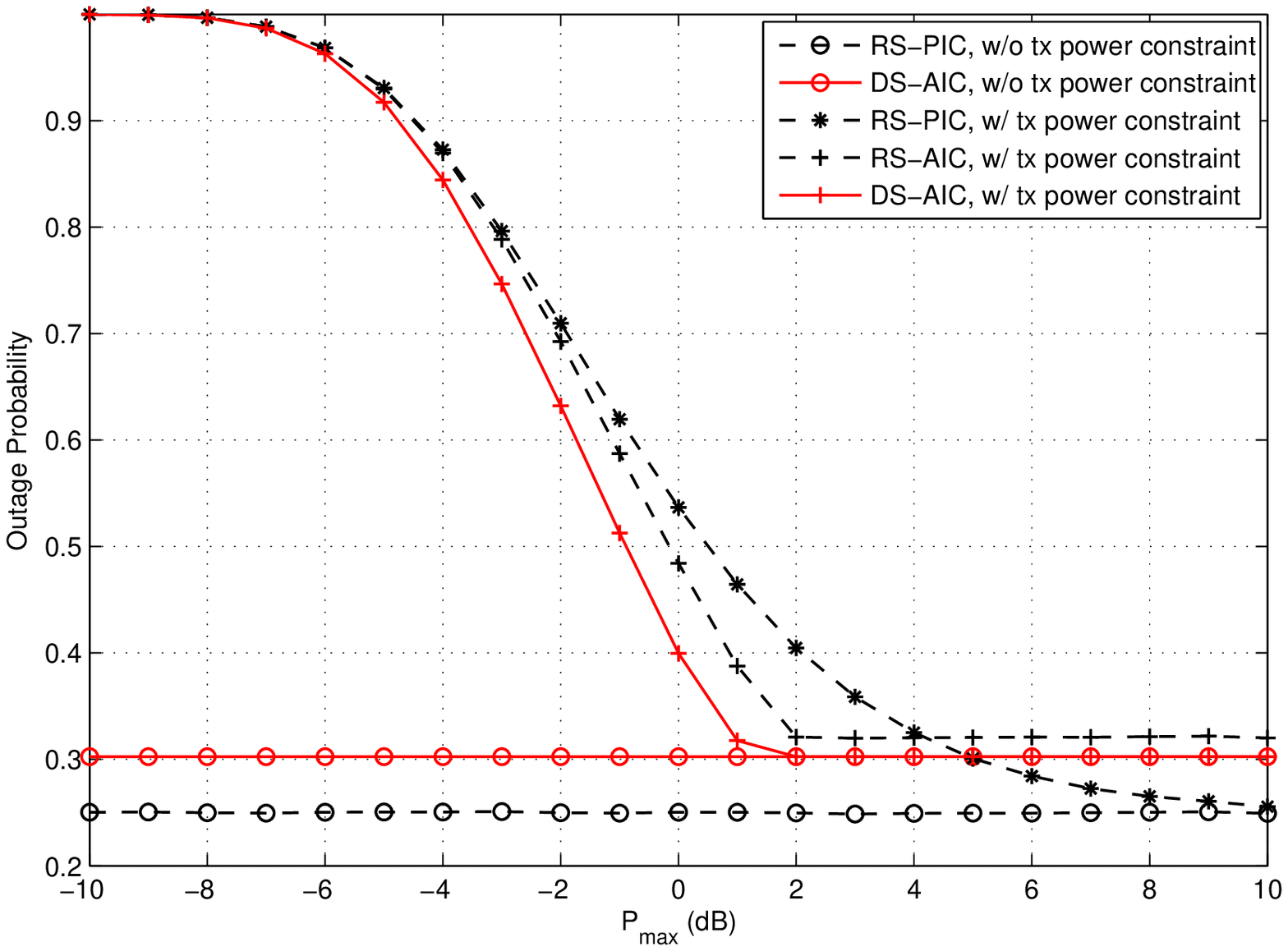}
     \caption{Comparison of outage probability between ratio selection and difference selection with various transmission and interference power constraints as a function of maximum allowable secondary transmission power $P_{\tiny{\mbox{max}}}$, where $M=2$ antennas at the secondary transmitter are used, and $\bar{\gamma}_s=\bar{\gamma}_p=1$. }
    \label{Pmax2}
\end{figure}

The results in the figures shown above apply when $P_{\tiny{\mbox{max}}}=0$ dB. In practice, the maximum allowable transmission power at the secondary transmitter can vary. To study the effect of the secondary transmission power constraint, we show in Figs. \ref{Pmax1} and \ref{Pmax2} mutual information and outage probability of the aforementioned five systems, where in all simulations it is assumed that $\bar{\gamma}_s=\bar{\gamma}_p=1$. It is observed from Fig.~\ref{Pmax1} that, when the impractical case is considered where no peak transmission power constraint is applied, RS-PIC outperforms all other systems as it is optimal in such a case. When this practical constraint is applied, however, the secondary link employing DS-AIC outperforms that employing RS-AIC in all the range of $P_{\tiny{\mbox{max}}}$ considered, and outperforms RS-PIC up to about $P_{\tiny{\mbox{max}}}=4$ dB. At a high $P_{\tiny{\mbox{max}}}$, a secondary link with RS-PIC outperforms that with RS-AIC and DS-AIC. Similar observations can be found in the outage probability plots shown in Fig.~\ref{Pmax2}. The reason that a secondary link employing RS-PIC outperforms DS-AIC when $P_{\tiny{\mbox{max}}}$ is large is because when $P_{\tiny{\mbox{max}}}$ is sufficiently high, the secondary transmission constraint is inactive. As a result, the secondary transmission power is only determined by the peak interference power, where in such a case RS-PIC is shown to be optimal \cite{wang2010capacity}. When $P_{\tiny{\mbox{max}}}$ is sufficiently low such that the interference constraints are inactive, the secondary transmission power of all the systems is limited by the same $P_{\tiny{\mbox{max}}}$, and they achieve the same mutual information and outage probability. In the medium range of $P_{\tiny{\mbox{max}}}$, a secondary link with AIC outperforms these with PIC is because AIC is a more relaxed constraint compared to PIC. 

\begin{figure}[t]
     \centering
     \includegraphics[width=\textwidth]{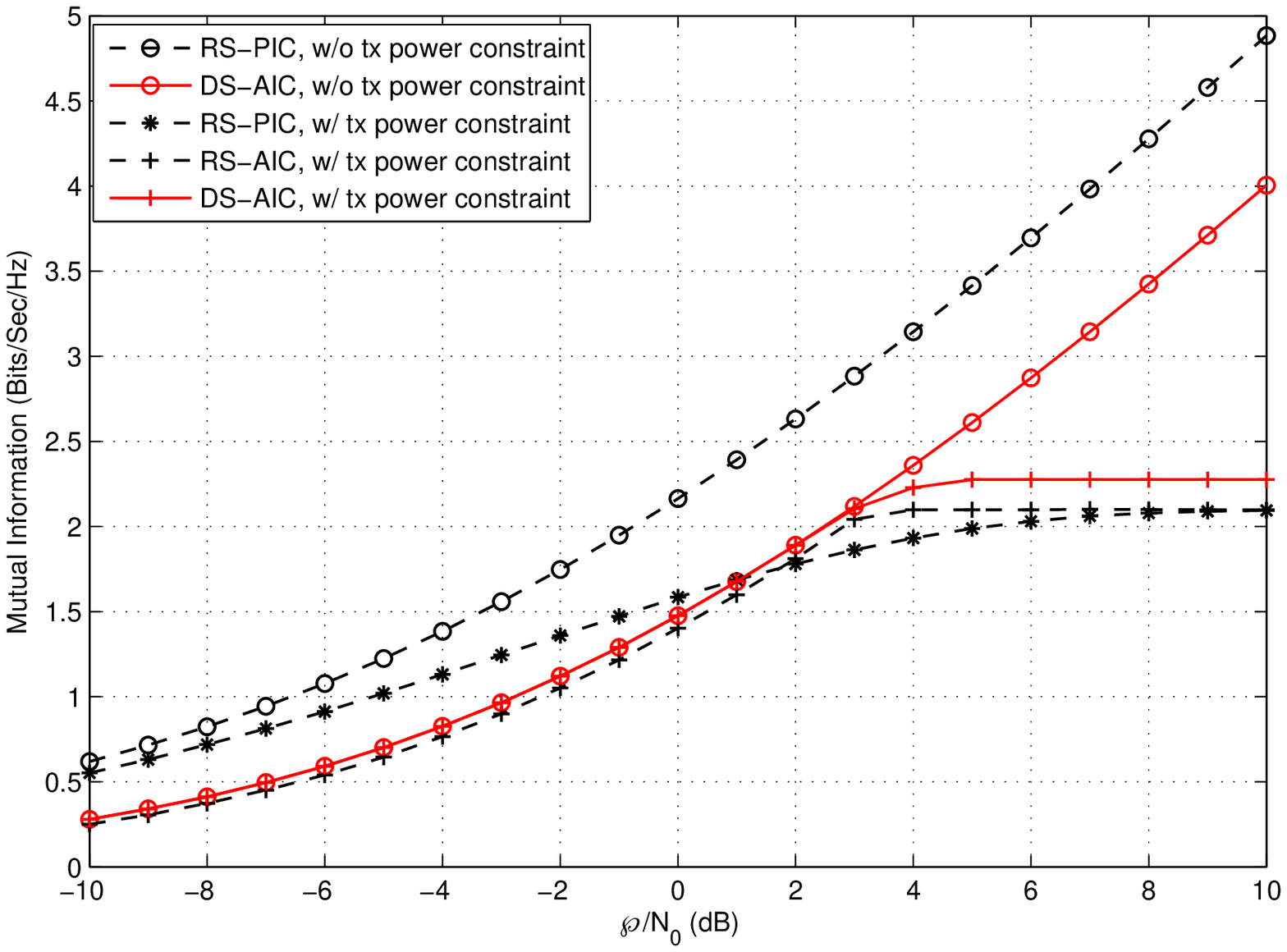}
     \caption{Comparison of mutual information between ratio selection and difference selection with various transmission and interference power constraints as a function of the interference threshold $\wp$, where $M=2$ antennas at the secondary transmitter are used, $\bar{\gamma}_s=\bar{\gamma}_p=1$, and $P_{\tiny{\mbox{max}}}=5$ dB.}
    \label{threshold1}
\end{figure}

\begin{figure}[t]
     \centering
     \includegraphics[width=\textwidth]{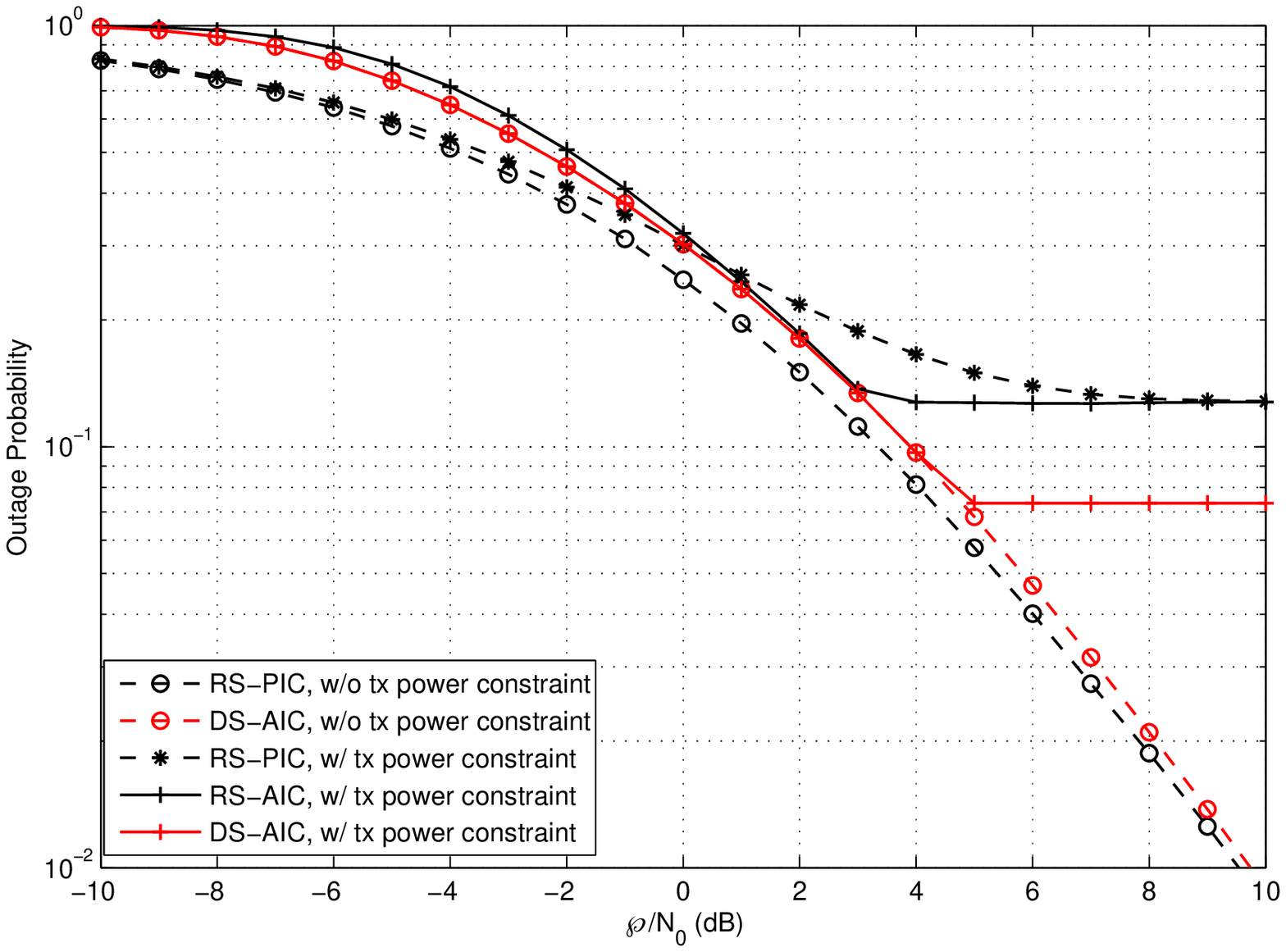}
     \caption{Comparison of outage probability between ratio selection and difference selection with various transmission and interference power constraints as a function of the interference threshold $\wp$, where $M=2$ antennas at the secondary transmitter are used, $\bar{\gamma}_s=\bar{\gamma}_p=1$, and $P_{\tiny{\mbox{max}}}=5$ dB.}
    \label{threshold2}
\end{figure}

Figs.~\ref{threshold1} and \ref{threshold2} compare the five aforementioned systems in terms of mutual information and outage probability with varying interference power constraints. It is assumed that $P_{\tiny{\mbox{max}}}=5$ dB when a secondary transmission power constraint is applied, and $\bar{\gamma}_s=\bar{\gamma}_p=1$. Improved mutual information and outage probability can be observed for all five systems as the interference power threshold increases, where RS-PIC without a secondary transmission power constraint outperforms all other systems. Again we stress here that a secondary system without a transmission power constraint is not practical, therefore the superiority of RS-PIC in such a case will not be beneficial to the secondary system in practice. When a peak secondary transmission power constraint is applied, RS-PIC achieves an improved mutual information and outage probability compared to systems with RS-AIC and DS-AIC when the interference power threshold is low. This is because in such a case, the secondary transmission power constraint is likely to be inactive, and the secondary transmission power is determined by the instantaneous channel gains of the $s \rightarrow p$ link when power allocation according to instantaneous channel knowledge is considered while it is determined by the average channel gains of the $s\rightarrow p$ link when power allocation according to channel statistics is considered. Since the instantaneous $s\rightarrow p$ channel gains are usually small due to the nature of ratio selection\footnote{This can be verified through simulation.}, a higher secondary transmission power is allowed for RS-PIC. When the interference power threshold becomes large and the secondary transmission power is by most cases determined by the maximum allowable transmission power, the advantage of RS-PIC diminishes, and its mutual information and outage probability degrade considerably. When the interference power threshold is sufficiently high such that the interference power constraint becomes inactive, RS-PIC and RS-AIC both achieve the same mutual information and outage probability because they specify the same secondary transmission power. However, they both yield an inferior performance compared to a secondary link with DS-AIC.

\section{Conclusion}\label{conclusion}
In this paper, difference antenna selection at the secondary transmitter in a wireless CR system has been proposed, where mutual information and outage probability of the secondary link due to such a selection method has been studied. Based on the analysis of mutual information and outage probability, a method of optimizing these performance metrics for the secondary data link subject to practical constraints on the peak secondary transmit power and the average interference power has been proposed, where the optimal selection weight $\delta$ and the transmission power allocation according to the channel statistics has been jointly determined. Comparisons between ratio selection and difference selection with various interference power constraints have shown that difference selection using the optimized parameters can be, in many cases of interest, superior to ratio selection.  

\section{Acknowledgment} 
The authors would like to thank the directors at TRL for their continued support. The authors would also like to thank Dr. Woon Hau Chen and Dr. Zhong Fan at TRL for the discussions at the initial stages of the work.

\appendices
\section{}\label{appendix1}
We give the proof of Lemma 1 in this appendix. For simplicity, we first introduce the following notations: $X_i=\gamma_{s,i}$, $Y_i=\gamma_{p,i}$, $\hat{X}=X_{\hat{\imath}}$ and $\hat{Y}=Y_{\hat{i}}$, where $\hat{\imath}$ is the index of the antenna that is selected due to difference antenna selection such that $Z_{\hat{\imath}}=\delta\gamma_{s,\hat{\imath}}-(1-\delta)\gamma_{p,\hat{\imath}}$ is maximized.  

\begin{proof}
The c.d.f. of $\hat{X}$ is given by
\begin{equation}\label{Fhatx}
F_{\hat{X}}(x)= P\left(\hat{X}\leq x\right) %&=& P\left\{\bigcup_{i=1}^M\left(\left(X_i\leq x\right)\bigcap_{j=1\atop i\neq j}^M\left(Z_i\geq Z_j\right)\right)\right\} \nonumber \\
%&=&M P\left\{\left(X_1\leq x\right)\bigcap\left(Z_1\geq Z_2\right)\bigcap\left(Z_1\geq Z_3\right)\cdots \bigcap\left(Z_1\geq Z_M\right)\right\}\nonumber \\
= M P\left(Z_1\geq \max\{Z_2,\cdots,Z_M\}\big| X_1 \leq x\right)P\left(X_1 \leq x\right).
\end{equation}
Denote $\breve{Z}=\max\left\{Z_2,\cdots, Z_M\right\}$, and $\breve{P}=P\left(Z_1\geq \breve{Z}\big| X_1 \leq x\right)$. Obtaining $\breve{P}$ requires the conditional p.d.f. of $Z_1$ and $\breve{Z}$ given $X_1\leq x$. 

The c.d.f. of $Z_1$ conditioned on $X_1\leq x$ is given by
\begin{equation}\label{P1}
F_{Z_1|X_1\leq x}(z)= \frac{P\left(Z_1\leq z, X_1 \leq x\right)}{P\left(X_1 \leq x\right)} = \frac{P\left(Y_1\geq \frac{\delta X_1-z}{1-\delta},X_1\leq x\right)}{P\left(X_1 \leq x\right)}
%&=& \left\{\begin{array}{ccc} I_1/F_{X_1}(x) &,& z>\delta x \\
%                              I_2/F_{X_1}(x) &,& 0 <z \leq \delta x \\
%                              I_3/F_{X_1}(x) &,& z \leq 0 \end{array}\right.
\end{equation}
%where $I_1$, $I_2$ and $I_3$ are due to different integral regions of $X_1$ and $Y_1$, given by 
%\begin{equation}
%I_1=\int_{0}^{x}\int_{0}^{+\infty}f_{X_1}(x_1)f_{Y_1}(y_1)dx_1dy_1 =1- \exp\left(-\frac{x}{\bar{\gamma}_s}\right)
%\end{equation}
%\begin{eqnarray}
%I_2 &=& \int_{0}^{z/\delta}\int_{0}^{+\infty}f_{X_1}(x_1)f_{Y_1}(y_1)dx_1dy_1 + \int_{z/\delta}^{x}\int_{\frac{\delta x_1-z}{1-\delta}}^{+\infty}f_{X_1}(x_1)f_{Y_1}(y_1)dx_1dy_1 \nonumber \\
%&=& 1-\frac{\delta\bar{\gamma}_s}{\bar{\gamma}}\exp\left(-\frac{z}{\delta\bar{\gamma}_s}\right)-\frac{(1-\delta)\bar{\gamma}_p}{\bar{\gamma}}\exp\left(\frac{z}{(1-\delta)\bar{\gamma}_p}\right)\exp\left(-\frac{\bar{\gamma}x}{\bar{\gamma}_s(1-\delta)\bar{\gamma}_p}\right)\nonumber \\
%\end{eqnarray}
%and
%\begin{eqnarray}
%I_3 &=& \int_{0}^{x}\int_{\frac{\delta x_1-z}{1-\delta}}^{+\infty}f_{X_1}(x_1)f_{Y_1}(y_1)dx_1dy_1 \nonumber \\
%&=& \frac{(1-\delta)\bar{\gamma}_p}{\bar{\gamma}}\exp\left(\frac{z}{(1-\delta)\bar{\gamma}_p}\right)\left(1-\exp\left(-\frac{\bar{\gamma}}{\bar{\gamma}_s(1-\delta)\bar{\gamma}_p}x\right)\right).
%\end{eqnarray}
which can be computed by considering different intervals of $z$ and integrating the joint p.d.f. of $X_1$ and $Y_1$ over the appropriate regions. Taking the derivative of $F_{Z_1|X_1\leq x}(z)$ over $z$ yields the p.d.f. of $Z_1$ given $X_1\leq x$, which is given by  
\begin{equation}\label{fz1x1}
f_{Z_1|X_1\leq x}(z) = \left\{\begin{array}{ccc} 0 &,& z>\delta x \\
                              \eta(x)\left(\exp\left(-\frac{z}{\delta \bar{\gamma}_s}\right)-\exp\left(-\frac{\bar{\gamma}}{\bar{\gamma}_s(1-\delta)\bar{\gamma}_p}x\right)\right)\exp\left(\frac{z}{(1-\delta)\bar{\gamma}_p}\right) &,& 0 <z \leq \delta x \\ 
                              \eta(x)\left(1-\exp\left(-\frac{\bar{\gamma}}{\bar{\gamma}_s(1-\delta)\bar{\gamma}_p}x\right)\right)\exp\left(\frac{z}{(1-\delta)\bar{\gamma}_p}\right) &,& z \leq 0 \end{array}\right.
\end{equation}
where $\bar{\gamma}=\delta\bar{\gamma}_s+(1-\delta)\bar{\gamma}_p$, and $\eta(x)=\frac{1}{F_{x_1}(x)}=\frac{1}{\bar{\gamma}\left(1-\exp\left(-\frac{x}{\bar{\gamma}_s}\right)\right)}$.

We now derive the conditional p.d.f. of $\breve{Z}$ given $X_1\leq x$. Following the definition of $Z_i$ in (\ref{Zi}), the p.d.f. of $Z_i$ is given by
\begin{equation}\label{zi}
f_{Z_i}(z) = \left\{\begin{array}{ccc} \frac{1}{\bar{\gamma}}\exp\left(\frac{z}{(1-\delta)\bar{\gamma}_{p}}\right)& , &z\leq 0 \\
\frac{1}{\bar{\gamma}}\exp\left(-\frac{z}{\delta\bar{\gamma}_{s}}\right)&, & z > 0   \end{array} \right..
\end{equation}
Since $\breve{Z}$ is independent of $X_1$, the conditional p.d.f. of $\breve{Z}$ given $X_1\leq x$ is the p.d.f. of $\breve{Z}$, which can be obtained by using the theory of ordering statistics \cite{david2003order}, giving
\begin{equation}\label{fbrevezx}
f_{\breve{Z}|X_1\leq x}(z) = f_{\breve{Z}}(z) = \left\{\begin{array}{ccc} \frac{(M-1)(1-\delta)^{M-2}\bar{\gamma}_{p}^{M-2}}{\bar{\gamma}^{M-1}}\exp\left(\frac{M-1}{(1-\delta)\bar{\gamma}_{p}}z\right)&,& z\leq 0 \\
\sum_{k=0}^{M-1}\left({M-1}\atop k\right)\frac{k\left(-\delta\bar{\gamma}_{s}\right)^{k-1}}{\bar{\gamma}^{k}}\exp\left(-\frac{k}{\delta\bar{\gamma}_{s}}z\right)&,& z> 0   \end{array} \right..
\end{equation}

Having (\ref{fz1x1}) and (\ref{fbrevezx}), we now derive $\breve{P}(x)=P\left(\breve{Z}-Z_1\leq 0|X_1\leq x\right)$. Since $Z_1$ and $\breve{Z}$ are independent, one has
\begin{equation}\label{Px}
\breve{P}(x)= \int_{-\infty}^{\delta x}\int_{-\infty}^{z_1}f_{Z_1|X_1\leq x}(z_1)f_{\breve{Z}}(z_2)dz_2dz_1 .
\end{equation}
Calculating the integral in (\ref{Px}) and substituting it into (\ref{Fhatx}) yields the result stated in Lemma 1. 
\end{proof}

\section{}\label{appendix4} 
We give the proof of Lemma 2 in this appendix. 

\begin{proof} 
Following the same notations as these given in Appendix I. The c.d.f. of the $s\rightarrow p$ link channel gain can be obtained in a similar manner as that for the $s\rightarrow s$ link, which is given by
\begin{equation}
F_{\hat{Y}}(y) = P\left(\hat{Y}\leq y\right) = MP\left(Z_1\geq \breve{Z}\big|Y_1\leq y\right)P(Y_1\leq y).
\end{equation}
Without going through the details of the derivation, we give the c.d.f. of $\hat{Y}$ as follows
\begin{equation}
F_{\hat{Y}}(y)= \left\{\begin{array}{ccc}
1-e^{-\frac{(1-\delta)\bar{\gamma}}{\varpi_s\varpi_p}y}+\frac{\varpi_p^{M}}{((M-1)\varpi_s-\varpi_p)\bar{\gamma}^{M-1}}\left(e^{-\frac{M}{\bar{\gamma}_p}y}-e^{-\frac{(1-\delta)\bar{\gamma}}{\varpi_s\varpi_p}y}\right)&,& (M-1)\varpi_s\neq \varpi_p \\
1-e^{-\frac{(1-\delta)\bar{\gamma}}{\varpi_s\varpi_p}y}-\frac{\varpi_p^{M}}{\bar{\gamma}^M}\left(e^{-\frac{M}{\bar{\gamma}_p}y}-e^{-\frac{(1-\delta)\bar{\gamma}}{\varpi_s\varpi_p}y}\right)-\frac{M(1-\delta)\varpi_p^{M-1}}{\bar{\gamma}^M}e^{-\frac{(1-\delta)\bar{\gamma}}{\varpi_s\varpi_p}y}y&,& (M-1)\varpi_s= \varpi_p
\end{array}\right..
\end{equation}
The p.d.f. of $\hat{Y}$, denoted as $f_{\hat{Y}}(y)$, can be obtained by taking the derivative of $F_{\hat{Y}}(y)$, and the expectation of $\mathbb{E}(\hat{Y})$ is given by
\begin{equation}\label{Ey}
\mathbb{E}(\hat{Y}) = \int_0^\infty yf_{\hat{Y}}(y)dy = \frac{M(\varpi_s+\varpi_p)^{M-1}\varpi_s+\varpi_p^M}{M(\varpi_s+\varpi_p)^M}\bar{\gamma}_p.\nonumber
\end{equation}
\end{proof}

\bibliographystyle{IEEEtran}
\bibliography{IEEEabrv,CR}

\newpage

%\begin{figure}[t]
%     \centering
%     \includegraphics[width=\textwidth]{max_ergodic_capacity_Pmax.eps}
%     \caption{Mutual Information}
%    \label{mutualinf3}
%\end{figure}
%
%\begin{figure}[t]
%     \centering
%     \includegraphics[width=\textwidth]{max_ergodic_capacity_threshold.eps}
%     \caption{Mutual Information}
%    \label{mutualinf4}
%\end{figure}

\end{document}